\documentclass[twocolumn,showpacs,preprintnumbers,amsmath,amssymb]{revtex4}
\usepackage{graphicx}
\usepackage{epsfig}
\usepackage{dcolumn}
\usepackage{bm}
\begin{document} 
\title{Evolving Networks with Multi-species Nodes \\ 
and Spread in the Number of Initial Links}
\author{Jong-Won Kim$^1$, Brian Hunt$^2$, and Edward Ott$^{1,3}$}
\affiliation{$^1$Department of Physics, and Institute for Research in 
Electronics and Applied Physics, \\
University of Maryland, College Park, Maryland  20742 \\ 
$^2$Department of Mathematics, and Institute for Physical Science and
Technology, \\
University of Maryland, College Park, Maryland  20742 \\ 
$^3$Department of Electrical and Computer Engineering, 
University of Maryland, College Park, Maryland  20742} 
\date{\today} 
 
\begin{abstract}
  We consider models for growing networks incorporating two effects not
previously considered: (i) different species of nodes, with each
species having different properties (such as different attachment probabilities 
to other node species); and (ii) when a new node is born, its number of links to
old nodes is random with a given probability distribution. Our numerical 
simulations show good agreement with analytic solutions.
As an application of our model, we investigate the movie-actor network with 
movies considered as nodes and actors as links. 
\end{abstract}
\maketitle 

\section{Introduction}

  It is known that many evolving network systems, including the world wide web, 
as well as social, biological, and communication systems, show power law 
distributions. In particular, the number of nodes with $k$ links is often 
observed to be $n_k \sim k^{-\nu}$, where $\nu$ typically varies from 2.0 to 3.1
\cite{Dorogovtsev1}. The mechanism for power-law network scaling was addressed 
in a seminal paper by 
Barab\'{a}si and Albert (BA) who proposed \cite{Barabasi1} 
a simple growing network model in which the probability of a new 
node forming a link with an old node (the ``attachment probability") is 
proportional to the number of links of the old node. This model yields a power 
law distribution of links with exponent $\nu = 3$. 
Many other works have been done extending this the model. 
For example Krapivsky and Redner \cite{Krapivsky1} provide a comprehensive 
description for a model with more general dependence of the attachment 
probability on the number $k$ of old node links. 
For attachment probability proportional to $A_k = a k + b$ they found that, 
depending on $b/a$,
the exponent $\nu$ can vary from 2 to $\infty$. Furthermore, 
for $A_k \sim k^\alpha$, when $\alpha < 1$, $n_k$ decays faster than a power 
law, while when $\alpha > 1$, there emerges a single node which connects to 
nearly all other nodes. Other modifications of the model are the introduction
of aging of nodes \cite{Dorogovtsev2}, initial attractiveness of nodes 
\cite{Dorogovtsev3}, the addition or re-wiring of links \cite{Albert1}, the
assignment of weights to links \cite{Yook1}, etc. 

  We have attempted to construct more general growing network models featuring
two effects which have not been considered previously: (i) multiple species of
nodes [in real network systems, there may be different species of nodes with 
each species having different properties (\it e.g.\rm, each species may have 
different 
probabilities for adding new nodes and may also have different attachment 
probabilities to the same node species and to other node species, etc.)]. 
(ii) initial link distributions [\it i.e.\rm, when a new node is born, its 
number of links 
to old nodes is not necessarily a constant number, but, rather, is 
characterized by a given probability distribution $p_k$ of new links]. 

  As an application of our model, we investigate the movie-actor network
with movies considered as nodes and actors as links (\it i.e.\rm, if the same 
actor appears in two movies there is a link between the two 
movies \cite{movie}). Moreover, we consider theatrical movies and 
made-for-television movies to constitute two different species. 

\section{Model}

  We construct a growing network model which incorporates multiple species and 
initial link probabilities. Given an initial network, we create new nodes at
a constant rate. We let the new node belong to species $j$ with 
probability $Q^{(j)}$ ($\sum_j Q^{(j)}=1$). We decide 
how many links $l$ the new node establishes with already existing nodes by 
randomly choosing $l$ from a probability distribution $p^{(j)}_l$. 
Then, we randomly attach the new node to $l$ existing nodes with preferential 
attachment probability proportional to a factor $A^{(j,i)}_k$, where $k$ is the
number of links of the target node of species $i$ to which the new node of 
species $j$ may connect. That is, the connection probability between an 
existing node and a new node is determined by the number of links of the 
existing node and the species of the new node and the target node.

  As for the single species case \cite{Krapivsky1}, the evolution of this model 
can be described by rate equations. In our case the rate equations give the
evolution of $N^{(i)}_k$, the number of species $i$ nodes that have 
$k$ links, 
\begin{eqnarray}
\frac{dN^{(i)}_k}{dt} &=& \sum^S_{j=1}Q^{(j)}\bar{k}^{(j)}
\frac{\left[A^{(j,i)}_{k-1} N^{(i)}_{k-1} - A^{(j,i)}_k N^{(i)}_k \right]}
{\sum_m \sum_k A^{(j,m)}_k N^{(m)}_k} \nonumber \\
& & + Q^{(i)}p^{(i)}_k,
\label{eq:Nk}
\end{eqnarray}
where $S$ is the total number of species and $\bar{k}^{(j)}=\sum_l l p^{(j)}_l$ 
is the average number of new links to a new node of species $j$, and $t$ is 
normalized so that the rate of creation of new nodes is one per unit time. 
The term proportional to 
$A^{(j,i)}_{k-1}N^{(i)}_{k-1}$ accounts for the increase of $N^{(i)}_k$ 
due to the addition of a new node of species $j$ that links to a species 
$i$ node with $k-1$ connections. The term proportional to 
$A^{(j,i)}_k N^{(i)}_k$ accounts for the decrease of $N^{(i)}_k$ due to 
linking of a new species $j$ node with an existing species $i$ node with 
$k$ connections. The denominator, $\sum_m \sum_k A^{(j,m)}_k N^{(m)}_k$,
is a normalization factor. If we add a new node with $l$ initial links, we have 
$l$ chances of increasing/decreasing $N^{(i)}_k$. This is accounted for by the 
factor $\bar{k}^{(j)} = \sum_l l p^{(j)}_l$ appearing in the summand of 
Eq. (\ref{eq:Nk}). The last term, $Q^{(i)}p^{(i)}_k$, accounts for the 
introduction of new nodes of species $i$. Since all nodes have at least one 
link, $N^{(i)}_0 = 0 $.

\section{Analysis of the Model} 

  Equation (\ref{eq:Nk}) implies that total number of nodes and total number 
of links increase at fixed rates. The total number of nodes of species $i$ 
increases at the rate $Q^{(i)}$. Thus
\begin{equation}
\sum_kN^{(i)}_k = Q^{(i)}t.
\label{eq:Qi}
\end{equation}
The link summation over all species $\sum_i \sum_k kN^{(i)}_k$ is twice the 
total number of links in the network. Thus
\begin{equation}
\sum^S_i \sum_k k N^{(i)}_k = 2 \left<\dot{k}\right>t,
\label{eq:kbar}
\end{equation}
where 
$\left<\dot{k}\right>=\sum_i\sum_kQ^{(i)}kp^{(i)}_k=\sum_iQ^{(i)}\bar{k}^{(i)}$.
Solutions of (\ref{eq:Nk}) occur in the form(\it c.f.\rm, \cite{Krapivsky1} for
the case of single species nodes),
\begin{equation}
N^{(i)}_k = n^{(i)}_k t,
\label{eq:nkt}
\end{equation}
where $n^{(i)}_k$ is independent of $t$. Eq. (\ref{eq:Nk}) yields 
\begin{equation}
n^{(i)}_k = \frac{B^{(i)}_{k-1}n^{(i)}_{k-1} + Q^{(i)}p^{(i)}_k}
{(B^{(i)}_k + 1)},
\label{eq:sol}
\end{equation}
where $B^{(i)}_k$ is
\begin{equation}
B^{(i)}_k = \sum^S_{j=1}Q^{(j)}\bar{k}^{(j)}
\frac{A^{(j,i)}_k} {\sum_m \sum_k A^{(j,m)}_k n^{(m)}_k}.
\label{eq:eta}
\end{equation}

\begin{figure}[t]
\epsfig{file=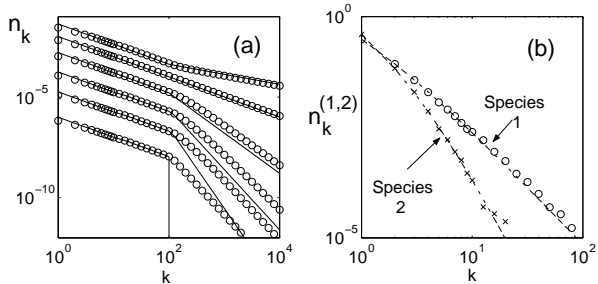, width=90mm}
\caption{(a) $n_k$ and $p_k$ versus $k$ for the single species network
model. Solid lines are the initial link probability $p_k$ and circles 
are the $n_k$ obtained from Eq. (\ref{eq:sol1}).
(b) $n^{(1)}_k$ and $n^{(2)}_k$ versus $k$ for the two species network model. 
Circles (species $1$) and crosses (species $2$) are log-binned data from our 
numerical simulation. The total number of nodes in our numerical network system 
is $10^6$. The dashed lines are solutions obtained from (\ref{eq:sol}) and
(\ref{eq:eta2}).}
\label{fig:model}
\end{figure}

  To most simply illustrate the effect of spread in the initial number of 
links, we first consider the case of a network with a single species of node
and with a simple form for the attachment $A_k = A^{(1,1)}_k$. In particular,
we choose \cite{Krapivsky1}, $A_k = k +c$. (Note that by Eq. (\ref{eq:Nk})
this is equivalent to $A_k = ak+b$ with $c=b/a$.)
Inserting this $A_k$ into Eq. (\ref{eq:eta}), we 
obtain $\sum_k (k+c)n_k = 2{\left<\dot{k}\right>} +cQ$ and  
$B_k = (k+c)/\eta$, where 
$\eta = (2{\left<\dot{k}\right>}+cQ)/{(Q\bar{k})} = 2 +c/{\bar{k}} \ge 2$. 
(Note that$\left<\dot{k}\right> = Q \bar{k}$ for the single species case.) 
Thus Eq. (\ref{eq:sol}) yields 
\begin{equation}
\left[(k+c) n_k - (k+c-1)n_{k-1} \right] + \eta n_k =  \eta Q p_k.
\label{eq:sol1}
\end{equation}
Setting $p_k = p_1 (k+c)^{-\beta}$, we can solve Eq. (\ref{eq:sol1}) for large 
$k$ by approximating the discrete variable $k$ as continuous, so that
\begin{equation}
(k+c) n_k - (k+c-1) n_{k-1} \cong \frac{d}{dk}[(k+c)n_k].
\label{eq:approx}
\end{equation}
Solution of the resulting differential equation,
\begin{equation}
\frac{d}{dk}[(k+c)n_k] + \eta n_k = \eta Q p_1 (k+c)^{-\beta},
\label{eq:sol2}
\end{equation}
for $n_k$ with $\beta \ne \eta+1$ consists of a homogeneous solution 
proportional to $(k+c)^{-(\eta+1)}$ plus the particular solution, 
$[\eta Q p_1/(\eta +1 -\beta)](k+c)^{-\beta}$. For $\beta = \eta+1$ 
the solution is $n_k = \eta Q p_1 (k+c)^{-(\eta +1)} \ln [d(k+c)]$, where $d$ is
an arbitrary constant. Hence, for \it sufficiently large \rm $k$ we have
$n_k \sim k^{-(\eta + 1)}$ if $\beta > \eta + 1$, and $n_k \sim k^{-\beta}$ 
if $\beta < \eta + 1$. Thus the result for $\beta > \eta + 1$ is independent of 
$\beta$ and, for $c = 0$, coincides 
with that give in Ref. \cite{Barabasi1} ($\eta +1 = 3$ when $c = 0$).
Solutions of Eq. (\ref{eq:sol1}) for $n_k$ versus $k$ in the range
$1 \le k \le 10^4$ are shown as open circles in Fig. \ref{fig:model}(a)
for initial link probabilities of the form
\begin{eqnarray}
p_k = \left\{ 
  \begin{array}{ll} 
    p_1 k^{-1} & \text{   for $1 \le k \le 10^2$} \\
    p_1 10^{2(\bar{\beta}-1)} k^{-\bar{\beta}} &\text{   for $k \ge 10^2$,}
  \end{array} 
\right.
\label{eq:pk}
\end{eqnarray}
which are plotted as solid lines in Fig. \ref{fig:model}(a).
The values of $\bar{\beta}$ used for the figure are 
$\bar{\beta} = 0.5, 1, 2, 3, 4$, and $\infty$ ($\bar{\beta}=\infty$ corresponds
to $p_k \equiv 0$ for $k > 10^2$). For clarity $n_k$ has been 
shifted by a constant factor so that $n_1$ coincides with the corresponding 
value of $p_1$. Also, to separate the graphs for easier visual inspection,
the value of $p_1$ for successive $\bar{\beta}$ values is changed
by a constant factor [since (\ref{eq:sol1}) is linear, the form of the solution
is not effected].
We note from Fig. \ref{fig:model}(a) that $n_k$ follows $p_k$ for $k < 10^2$
in all cases. This is as expected, since $p_k$ decreases slower than $k^{-3}$
in this range. Furthermore, $n_k$ very closely follows $p_k$ for $k > 10^2$
for $\bar{\beta} = 0.5, 1.0, 2.0$. As $\bar{\beta}$ increases deviations of 
$n_k$ from $p_k$ in $k > 10^2$ become more evident, and the large $k$ asymptotic
$k^{-3}$ dependence is observed. Thus, if $p_k$ decreases sufficiently rapidly,
then the behavior of $n_k$ is determined by the growing network dynamics, 
while, if $p_k$ decreases slowly, then the behavior of $n_k$ is determined
by $p_k$.

  To simply illustrate the effect of multiple species we now consider a 
growing two species network with $p_k = \delta_{1,k}$ (\it i.e.\rm,
$p_k = 0$ for $k \ge 2$). Then, Eq. (\ref{eq:eta}) becomes
\begin{subequations} \label{eq:eta1}
\begin{align}
B^{(1)}_k &=& 
\frac{Q^{(1)}A^{(1,1)}_k}{\sum_m \sum_k A^{(1,m)}_k n^{(m)}_k} + 
\frac{Q^{(2)}A^{(2,1)}_k}{\sum_m \sum_k A^{(2,m)}_k n^{(m)}_k}, 
\label{eq:eta1a} \\
B^{(2)}_k &=& 
\frac{Q^{(1)}A^{(1,2)}_k}{\sum_m \sum_k A^{(1,m)}_k n^{(m)}_k} + 
\frac{Q^{(2)}A^{(2,2)}_k}{\sum_m \sum_k A^{(2,m)}_k n^{(m)}_k},
\label{eq:eta1b}
\end{align}
\end{subequations}
where $\sum_m$ represents summation of species $1$ and $2$ nodes.

  In order to illustrate the model with our numerical simulations, we specialize
to a specific case. We choose attachment coefficients 
$A^{(1,1)}_k = ak$, $A^{(1,2)}_k = ak$, $A^{(2,1)}_k = bk$, and $A^{(2,2)}=0$.
Thus a new species $1$ node connects to either existing 
species $1$ nodes and species $2$ nodes with equal probability, while a new 
species $2$ node can connect to existing species $1$ nodes only.
Therefore, the first summation term in Eq. (\ref{eq:eta1}), 
${\sum_m \sum_k A^{(1,m)}_k n^{(m)}_k}$, becomes 
$a\sum_k(kn^{(1)}_k+kn^{(2)}_k)$,
which is $a$ times the total increase of links at each time 
$a \times 2(Q^{(1)}+Q^{(2)})$. Recall that 
$Q^{(1)} + Q^{(2)} = 1$. In order to calculate the second summation term in 
Eq. (\ref{eq:eta1}), 
${\sum_m \sum_k A^{(2,m)}_k n^{(m)}_k} = b \sum_k k n^{(1)}_k$, 
we define a parameter $\gamma$ that is the 
ratio of the total number of links of species $1$ to the total number of links 
in the network. Since the probability of linking a new species $1$ node to 
existing species $1$ nodes is determined by the total number of links of 
species $1$,
this probability is exactly same as $\gamma$. Thus, if we add a new species $1$ 
node, the number of links of species $1$ increases by $Q^{(1)}$ due to
the new node and by $\gamma Q^{(1)}$ due to the existing species $1$ nodes 
that become
connected with the new node, while the number of links of species $2$
increases by $(1-\gamma)Q^{(1)}$.
But, if we add a new species $2$ node, the numbers of links 
increases by $Q^{(2)}$ for both species because a new species $2$ node can 
link to species $1$ nodes only. Thus, the increase of species $1$ links is 
$(1+\gamma)Q^{(1)} + Q^{(2)}$ and that of species $2$ links is 
$(1-\gamma)Q^{(1)} + Q^{(2)}$. Since $\gamma$ is the ratio of the 
number of species $1$ links to the total number of links,
$\gamma = [(1+\gamma)Q^{(1)} + Q^{(2)}]/2$ or
\begin{equation}
\gamma = \frac{1}{2-Q^{(1)}}.
\label{eq:gamma}
\end{equation} 
With this $\gamma$, Eq. (\ref{eq:eta1}) becomes
\begin{subequations} \label{eq:eta2}
\begin{eqnarray}
B^{(1)}_k &=& \frac{Q^{(1)}}{2}k + \frac{Q^{(2)}(2-Q^{(1)})}{2}k 
= \frac{k}{\eta^{(1)}}, 
\label{eq:eta2a} \\
B^{(2)}_k &=& \frac{Q^{(1)}}{2}k = \frac{k}{\eta^{(2)}}. 
\label{eq:eta2b}
\end{eqnarray}
\end{subequations}
where obtain $\eta^{(1)} = 2/[Q^{(1)}+Q^{(2)}(2-Q^{(1)})]$ 
and $\eta^{(2)} = 2/Q^{(1)}$. 

  Proceeding as for the single species case, we approximate (\ref{eq:sol}) by
an ordinary differential equation (\it c.f.\rm, Eq. (\ref{eq:sol2})) to obtain
$n^{(i)}_k \sim k^{-(1+\eta^{(i)})}$. As an example, we set 
$Q^{(1)} = Q^{(2)} = 0.5$, in which case Eqs. (\ref{eq:eta2}) give exponents
$1+\eta^{(1)}=2.6$ and $1+\eta^{(2)} = 5$. In Fig. \ref{fig:model}(b) we plot,
for this case, the analytic solution obtained from (\ref{eq:sol}) and
(\ref{eq:eta2}) as dashed lines, and the results of numerical simulations
as open circles and pluses. The simulation results, obtained by histogram
binning with uniform bin size in $\log k$, agree with the analytic solutions,
and both show the expected large $k$ power law behaviors, 
$n^{(1)}_k \sim k^{-2.6}$ and $n^{(2)}_k \sim k^{-5}$. 

\section{The Movie-Actor Network}

  We now investigate the movie-actor network. We collected data from the 
Internet Movie Data Base (IMDB) web site \cite{Imdb}. The total number of 
movies is 285,297 and the total number of actors/actresses is 555,907. Within 
this database are 226,325 theatrical movies and 24,865 made for television 
movies. The other movies in the database are made for television series, 
video, mini series, and video games. In order to get good statistics,
we choose only theatrical and television movies made 
between 1950 to 2000. Thus we have two species of movies.
We also consider only actors/actresses from these movies.
We consider two movies to be linked if they have an actor/actress in common.
We label the theatrical movies species $1$, and the made for television
movies species $2$.

\begin{figure}[t]
\epsfig{file=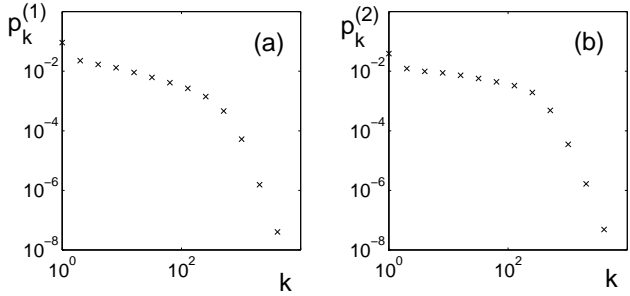,width= 8.5 cm}
\caption{The initial link probability distributions $p_k$ of
(a) theatrical movies and (b) television movies. These plots are obtained 
using bins of equal width in $\log k$ and dividing the number of nodes in
each bin by the product of the bin width in $k$ (which varies from bin to bin)
and the total number of nodes.}
\label{fig:pks}
\end{figure}

  In order to apply our model, Eq. (\ref{eq:Nk}), we require as input
$Q^{(j)}, p^{(j)}_k$ and $A^{(j,i)}_k$ which we obtain from the movie-actor
network data. We take $Q^{(1)}$ and $Q^{(2)}$ to be, respectively, 
the fractions of theatrical and made for television movies in our data base.
We obtain $Q^{(1)} = 0.83$ and $Q^{(2)} = 0.17$. We now consider $p^{(j)}_k$.
Suppose a new movie is produced casting $r$ actors. For each actor $s$
$(s = 1, 2, ..., r)$ let $l_s$ denote the number of previous movies in which
that actor appeared. Then the total number of the
initial links of the new movie is $\sum_s l_s$. From histograms of this 
number, we obtain (Figs. \ref{fig:pks}) the initial link probability 
distributions $p^{(j)}_k$. 

\begin{figure}[t]
\epsfig{file=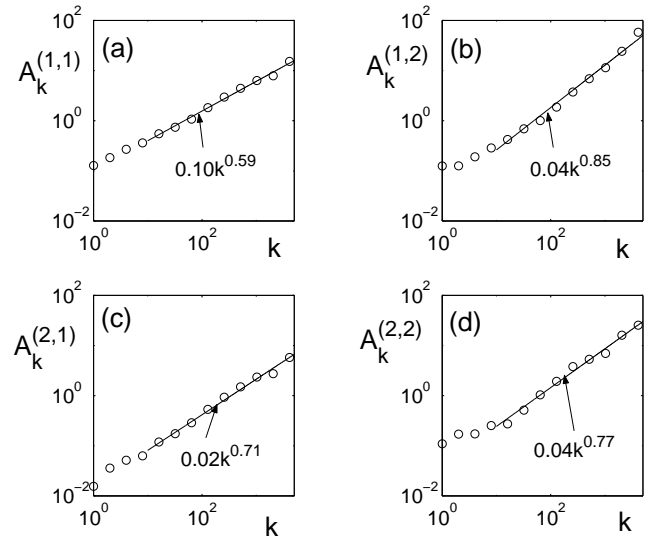,width= 8.5 cm}
\caption{Attachment coefficients for theatrical movies (a) $A^{(1,1)}_k$ and 
(b) $A^{(1,2)}_k$, and for television movies (c) $A^{(2,1)}_k$ and 
(d) $A^{(2,2)}_k$. All data are obtained using log-binning without 
normalization (see caption to Fig. \ref{fig:pks}).}
\label{fig:aks}
\end{figure}

The attachment $A^{(j,i)}_k$ can be numerically obtained from data via, 
\begin{equation}
A^{(j,i)}_k \sim \frac{\left<\Delta(j;i,k)\right>}{\delta t},
\label{eq:ak}
\end{equation}
where $\Delta(j;i,k)$ is the increase during a time interval $\delta t$  
in the number of links between old species $i$ nodes that had $k$ links
and new species $j$ nodes, and $<...>$ is an average over all such species
$i$ nodes \cite{Jeong1}. In the movie network, we count all movies and links 
from 1950 to 1999, and measure the increments in the number of links for a 
$\delta t$ of one year. We obtain attachment coefficient 
$A^{(1,1)}_k \sim 0.10k^{0.59}$ and 
$A^{(1,2)}_k \sim 0.04k^{0.85}$ for theatrical movies, 
and $A^{(2,1)}_k 0\sim 0.02k^{0.71}$ and $A^{(2,2)}_k \sim 0.04k^{0.77}$ 
for television movies. See Fig. \ref{fig:aks}.

  Incorporating these results for $Q^{(i)}$, $p^{(i)}_k$ and $A^{(j,i)}_k$
in our multi-species model, Eq. (\ref{eq:Nk}), we carry out numerical 
simulations as follows: 
(i) We add a new movie at each time step. We randomly designate each new movie
as a theatrical movie with probability $Q^{(1)}=0.83$ or a television movie 
with probability $Q^{(2)}=0.17$. 
(ii) With initial link probability $p^{(j)}_k$,
we randomly choose the number of connections to make to old movies.  
(iii) We then use the attachment $A^{(j,i)}_k$ to randomly choose connections
of new species $j$ movie to old species $i$ movies. 
(iv) We repeat (i)-(iii) adding 100,000 new movies, and finally calculate 
the probability distributions of movies with $k$ links. 
 
\begin{figure}[b]
\epsfig{file=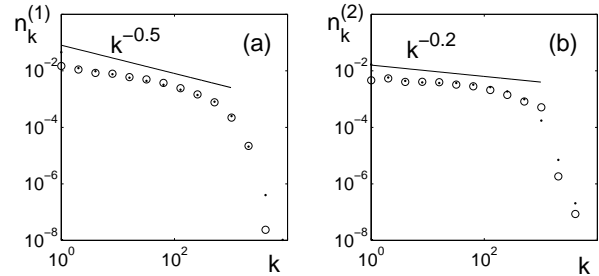,width= 8 cm}
\caption{The probability distributions $n^{(i)}_k$ of movies that have $k$
links; (a) theatrical movies $n^{(1)}_k$ and (b) television movies $n^{(2)}_k$.
Dots are $n^{(i)}_k$ obtained from the movie network while circles are 
from numerical simulation using $Q^{(j)}$ obtained from our data base, 
$p^{(j)}_k$ in Fig. \ref{fig:pks} and $A^{(j,i)}_k$ in Fig. \ref{fig:aks}. 
All data are obtained using log-binning (see caption to Fig. \ref{fig:pks}).}
\label{fig:nk}
\end{figure}

\begin{figure}[t]
\epsfig{file=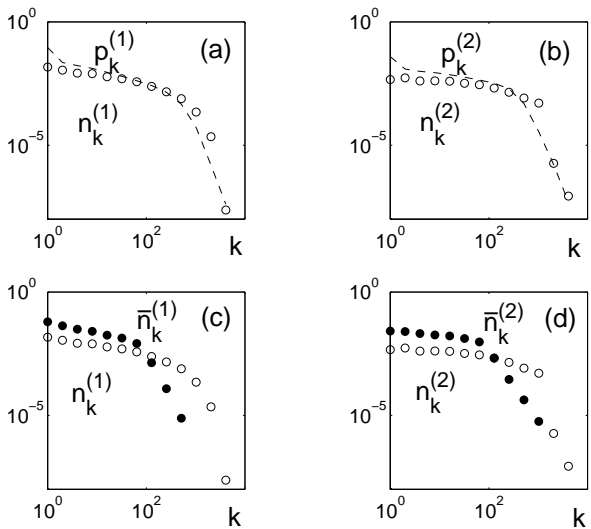,width= 8 cm}
\caption{(a) and (b) are $n^{(i)}_k$ (circles) obtained from numerical 
simulations using $p^{(i)}_k$ (dashed lines), while (c) and (d) show 
$n^{(i)}_k$ from (a) and (b) (open circles) plotted with results denoted
$\bar{n}^{(i)}_k$ (filled circles) from simulation using  a cutoff initial
link probability $\bar{p}^{(i)}_k$ 
(where $\bar{p}^{(i)}_k = p^{(i)}_k/\sum \bar{p}^{(i)}_k$ when
$k \le 50$ and $\bar{p}^{(i)}_k = 0$ when $k > 50$). 
All data are obtained using log-binning (see caption to Fig. \ref{fig:pks}).}
\label{fig:pknk}
\end{figure}

  Figure \ref{fig:nk} shows $n^{(i)}_k$ versus $k$ obtained from our 
movie-actor network data base (dots) and from numerical simulations using 
Eq.(\ref{eq:Nk}) (open circles) with our empirically obtained 
results for $Q^{(j)}$, $p^{(j)}_k$, and $A^{(j,i)}_k$. The results are 
are roughly consistent with the existence of two scaling regions 
\cite{twoscale1}.
For small $k$ $(k \lesssim 10^2$) the two species exhibit slow power law
decay with different exponents, $n^{(1)}_k \sim k^{-0.5}$, 
$n^{(2)}_k \sim k^{-0.2}$, while for large $k$ the probabilities decay
much more rapidly. Indeed, the results of \cite{Krapivsky1} suggest that
the decay should be exponential for large $k$ since the attachment
$A^{(j,i)}_k$ grow sub-linearly with $k$.
We showed in Sec. III for the single species 
model with a linear attachment $A_k \sim k$ that $n_k$ follows $p_k$ when 
$p_k$ decays slowly, while $n_k$ is independent of $p_k$ when $p_k$ decays 
sufficiently quickly. As we will later show, this feature is also applicable to 
multi-species networks with nonlinear attachments. 
As seen in Figs. \ref{fig:pknk}(a) and \ref{fig:pknk}(b), $n^{(i)}_k$ 
follows $p^{(i)}_k$ in the small $k$ region. However, it is not clear whether 
$n^{(i)}_k$ follows $p^{(i)}_k$ in the large $k$ region.
In order to check the behavior of $n^{(i)}_k$ in this region, we carried out 
another numerical simulation using an initial link probability 
$\bar{p}^{(i)}_k$ which is cut off at $k=50$. 
That is, $\bar{p}^{(i)}_k = p^{(i)}_k/\sum \bar{p}^{(i)}_k$ when 
$k \le 50$ and $\bar{p}^{(i)}_k = 0$ when $k > 50$. 
Using $\bar{p}^{(i)}_k$ in place of $p^{(i)}_k$, 
we obtain from our simulation corresponding data, $\bar{n}^{(i)}_k$ versus
$k$, which are shown in Figs. \ref{fig:pknk}(c) and \ref{fig:pknk}(d) as
filled in circles. For comparison the data for $n^{(i)}_k$ from Figs.
\ref{fig:pknk}(a) and \ref{fig:pknk}(b) are plotted in Figs. \ref{fig:pknk}(c)
and \ref{fig:pknk}(d) as open circles. It is seen that the cutoff at $k=50$
induces a substantial change in the distribution of the number of links
for $k>50$. Thus it appears that, in the range tested, the large $k$ behavior 
of the movie-actor network is determined by the initial link probability 
$p^{(i)}_k$ rather than by the dynamics of the growing network.


  In conclusion, in this paper we propose a model for a multi-species network
 with variable initial link probabilities. We have investigated the 
movie-actor network as an example. We believe that the effect of multiple 
species nodes may be important for modeling other complicated networks
(\it e.g.\rm, the world wide web can be divided into commercial sites and 
educational or personal sites).  We also conjecture that the initial link 
probability is a key feature of many growing networks.
 
\begin{references}
\bibitem{movie} Barab\'{a}si and Albert also investigated the movie-actor 
network. However, they consider actors as nodes that are linked if
they are cast in the same movie. See Ref. \cite{Barabasi1} and Ref.
\cite{Albert1}.
\end{references}
\end{document}